\documentclass[aps,prl,twocolumn,letter]{revtex4-1}

\usepackage[normalem]{ulem}

\usepackage{graphicx}  
\usepackage{bm}        
\usepackage{amssymb}   
\usepackage[usenames]{xcolor}
\usepackage{ulem}
\usepackage{bbding}
\usepackage{ifsym}
\usepackage{MnSymbol}
\usepackage{pifont}
\usepackage{latexsym}
\usepackage{wasysym}
\usepackage{subfig}
\usepackage{tabularx}
\newcolumntype{s}{>{\hsize=.5\hsize}X}

\begin{document}

\title{Degree product rule tempers explosive percolation in the absence of global information}

\date{\today}
\author{Alexander J. Trevelyan, Georgios Tsekenis, Eric I. Corwin}
\affiliation{Materials Science Institute and Department of Physics, University of Oregon, Eugene, Oregon 97403}

\begin{abstract}
We introduce a guided network growth model, which we call the degree product rule process, that uses solely local information when adding new edges. For small numbers of candidate edges our process gives rise to a second-order phase transition, but becomes first-order in the limit of global choice. We provide the set of critical exponents required to characterize the nature of this percolation transition. Such a process permits interventions which can delay the onset of percolation while tempering the explosiveness caused by cluster product rule processes.
\end{abstract}

\maketitle

\textit{Introduction.}--Network based approaches continue to see growing applications in a wide array of fields, from epidemiology \cite{disease,proteins} to finance \cite{finance1,finance2}, neuroscience \cite{brain1,brain2}, and machine learning \cite{machine}. As we increasingly rely on networks, understanding how they form out of complex conditions becomes all the more consequential \cite{StrogatzReview,NotreReview,NewmanReview,AveiroReview,EpidemicReview}. Many the networks we entrust to support our modernized society--transportation, financial, social, etc.--are formed with some amount of agency, meaning that potential new members have control over how they connect and interact with the network. This agency can lead to markedly different behavior compared to the classical case of purely random network growth \cite{DoyleAgency}. In particular, networks subject to competitive edge addition break time-reversal symmetry, as there is no well-defined method for running the process in reverse that achieves a statistically identical growth curve \cite{continuousEP}. Furthermore, edge competition can be used as a means of control over cluster growth and connectivity within a growing network. Depending on the desired outcome (delayed connectivity for contagion spreading, increased connectivity for communication networks, etc.), intervening on growing networks can help produce more specialized and responsive networks.

Pioneering work by Erd\H os and R\'{e}nyi \cite{ER} characterized the most straightforward process of random network growth: edges are added to the network uniformly at random until connectivity \textit{percolates} through the entire network. The Achlioptas growth process (AP) \cite{explosiveOG} adds a layer of competition to the classical percolation process, whereby edges are ranked based on the sizes of the clusters they join and then added to the network in such a way as to suppress large cluster growth. This competition results in a significant delay in the onset of percolation, but comes at the cost of a much more abrupt transition--it produces what is commonly referred to as ``powder keg'' conditions \cite{powderkeg1,powderkeg2}, where clusters in a narrow band of size become widespread and primed for sudden connectivity. The powder keg formation can be mitigated by continuously adding new nodes to the network \cite{newnodearrival1,newnodearrival2}, inducing an infinite-order transition, however in many real-world cases such an intervention is impractical.

Variations in competitive edge addition, such as the minimal cluster rule \cite{powderkeg1}, the triangle rule \cite{trianglerule}, and a handful of others covered in the review article in reference \cite{review1}, achieve results similar to the AP. Together, these growth processes are referred to as \textit{explosive percolation} due to the abruptness with which the largest cluster grows from microscopic to system-spanning. Each of these processes shares a common thread: edge competition involves comparing the sizes of the clusters to which each edge belongs, which necessitates gathering information about the connectivity of a large portion of the network as it nears the percolation threshold. Though generally second-order \cite{continuousEP}, under certain circumstances these transitions can become first-order \cite{discontinuousEP1}, typically when either the number of edges competing for addition at each timestep grows quickly enough with system size \cite{continuousEP2}, or the competition process is designed to build up smaller clusters that eventually merge together and overtake the largest component \cite{discontinuousEP2}. Approaches focused on local measures of connectivity \cite{degreemacro,localrules} have reproduced some aspects of explosive percolation, yet remain relatively unexplored compared to global product rules. Additional novel phenomena that have been observed in explosive percolation including crackling noise and ``fractional percolation'' \cite{crackling}, unexpected double-peaked distributions of the order parameter in small systems \cite{unusualFSS}, and finite-size hysteresis \cite{hysteresis}.

Here, we introduce and characterize the behavior of a third type of random growth process, the \textit{degree product rule} (DPR) process. Mechanistically, the DPR is analogous to the Achlioptas process, the difference being that the criteria used to evaluate edges is the product of node degrees (the number of edges attached to a node) rather than cluster sizes. The impetus for studying such a subtle but fundamental modification is twofold. First, the degree of a node is local information in the sense that for any given node, determining its degree requires only knowledge of its set of nearest neighbors. Unlike average cluster size, information about the average degree of each node does not become extensive within the system near the percolation threshold. Second, the problem of classical percolation has long involved using a stable probability distribution to choose an edge at each timestep. Explosive percolation upended this notion by allowing the distribution to shift unpredictably depending on which edge is chosen, a characteristic potentially more in line with how certain types of real networks take shape \cite{realworld}. The DPR similarly produces unpredictable changes when updating edge selection probabilities, but does so under a set of local rules, broadening our understanding of how networks coalesce under various formational pressures.

\textit{The degree product rule model.}--We begin with a fully disconnected set of $N$ nodes and successively add edges one at a time, such that at time $t$ the network contains exactly $t$ edges, with a resulting edge density $p = t/N$. The growth process is as follows: 1) A specified number of candidate edges $m$ are chosen uniformly at random. 2) The weight of each candidate edge is calculated as the product of the degrees $d$ of the two nodes to be connected by that edge as $(d_{1} + 1)(d_{2} + 1)$, where one is added to the degree of each node in order to avoid the degenerate case of zero-degree nodes. 3) The edge with the smallest weight is added to the network or, in the case of a tie, an edge is chosen at random from the set of edges with the smallest weight. The remaining edges are discarded back into the pool of unfilled edges. The process is illustrated diagrammatically in the inset of Figure 1.

During any random growth process clusters will form, grow, and eventually merge together. The relative size of the largest cluster $C/N$ is computed at every timestep and serves as the order parameter of the percolation transition. The order parameter begins vanishingly small, then becomes macroscopic as the system crosses the critical point $p_{c}$, the precise value of which is determined by the details of the growth process. Figure 1 shows the ensemble-averaged evolution of the order parameter for the Erd\H os-R\'{e}nyi, Achlioptas, and DPR processes, with $m = 2$ for the latter two. In principle, the critical point of each transition can be predicted by analyzing the combinatorics of the system, however in practice this becomes prohibitively difficult when the underlying distribution used to add edges changes unpredictably as in the AP and DPR. Thus, numerical simulations are necessary to tackle the details of these systems and obtain precise approximations of their critical behavior.

\begin{figure}
  \includegraphics[width=\linewidth]{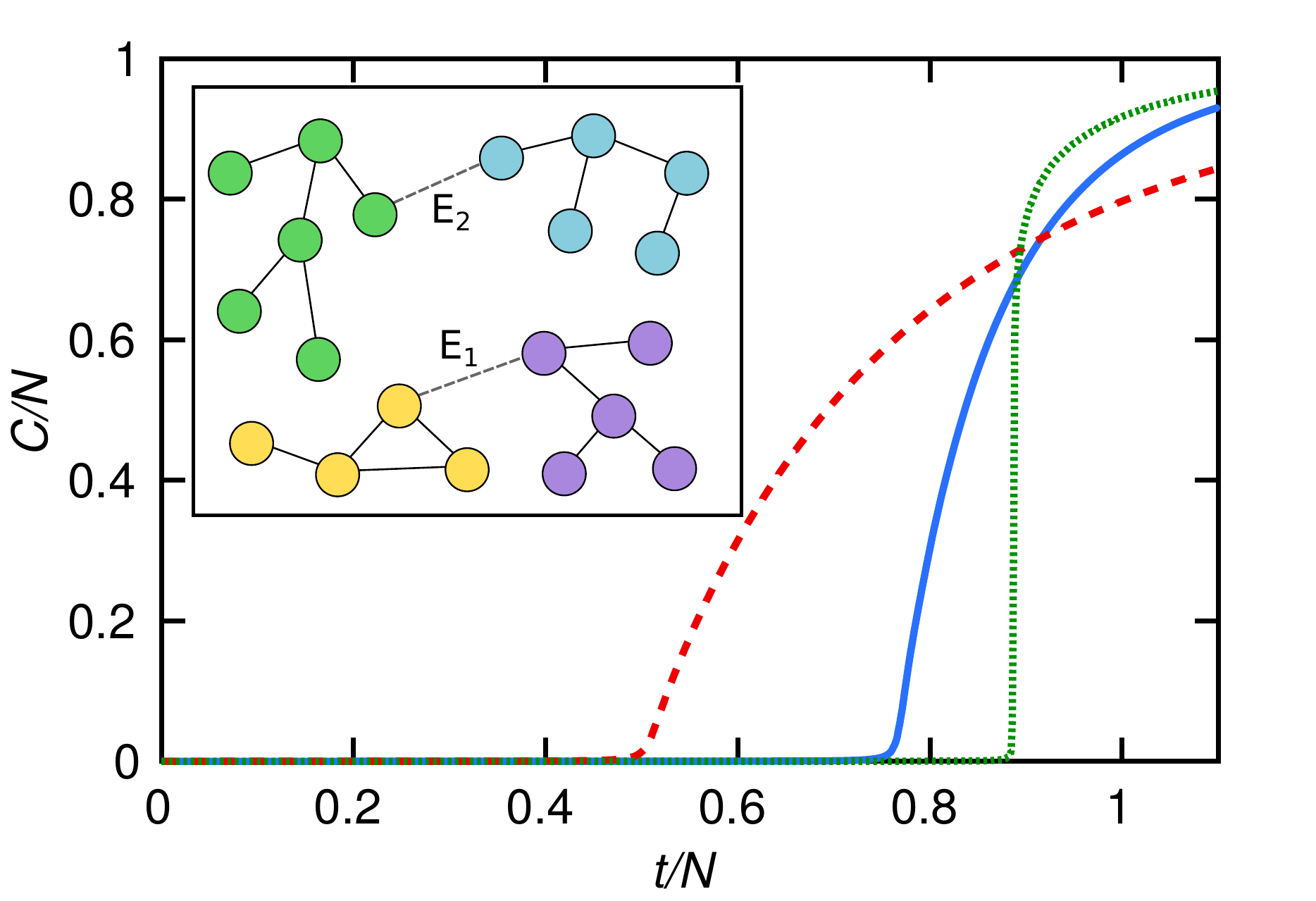}
  \caption{Relative size of the largest cluster, $C/N$, at scaled time $p = t/N$. Ensemble averages for Erd\H os-R\'{e}nyi (dashed red), DPR process (solid blue), and AP (dotted green) at $N = 3.6\times10^{5}$ nodes and $m = 2$ choices for the DPR process and AP. Inset: Example of the DPR selection scheme for $m = 2$ choices. $E_{1}$ and $E_{2}$ compete for addition. The selection criteria $A = (d_{1} + 1)(d_{2} + 1)$ is computed for each edge. Since $A_{E_{1}} = 9$ and $A_{E_{2}} = 4$, $E_{2}$ is added to the network.}
  \label{MeanRunsWithDiagramInset}
\end{figure}

The percolation transitions presented in Figure 1 are notably different in both the location of the critical point and abruptness of each transition. To better quantify the abruptness of the DPR transition, we measure the size of the largest jump in the order parameter $\Delta C_{max}/N$ during each realization, then average over many realizations. This type of convergence criteria is common among explosive percolation studies \cite{singlelink,review2,EPmultiplegiant,discontinuousEP3} as it gives insight into how the transition behaves in the thermodynamic limit and indicates whether the transition is first-order or second-order. For increasing system size, the largest jump will decay as a power law when the transition is second-order, $\Delta C_{max}/N \sim N^{-\omega}$, whereas if there is a discontinuity that survives in the thermodynamic limit then $\Delta C_{max}/N$ will asymptote to a constant value, signaling that the transition is first-order. The decay exponent $\omega$ communicates the level of the abruptness in second-order transitions, with smaller values indicating a sharper transition. In the AP, the decay exponent is unusually small: $\omega = 0.065$ for $m = 2$ choices. The DPR, however, produces decay exponents similar to Erd\H os-R\'{e}nyi, as shown in Figure 2. In fact, despite the appearance of a faster transition, the DPR is actually seen to have a decay exponent only slightly larger than Erd\H os-R\'{e}nyi, recorded in Table I. In addition, finite-size effects show up at small system sizes for the DPR between $N = 10^{2}$ up to $N = 10^{4}$ in Figure 2, depending on the number of choices,  whereas in both Erd\H os-R\'{e}nyi and explosive percolation no such effects appear at comparable system sizes. 

\begin{figure}
  \includegraphics[width=\linewidth]{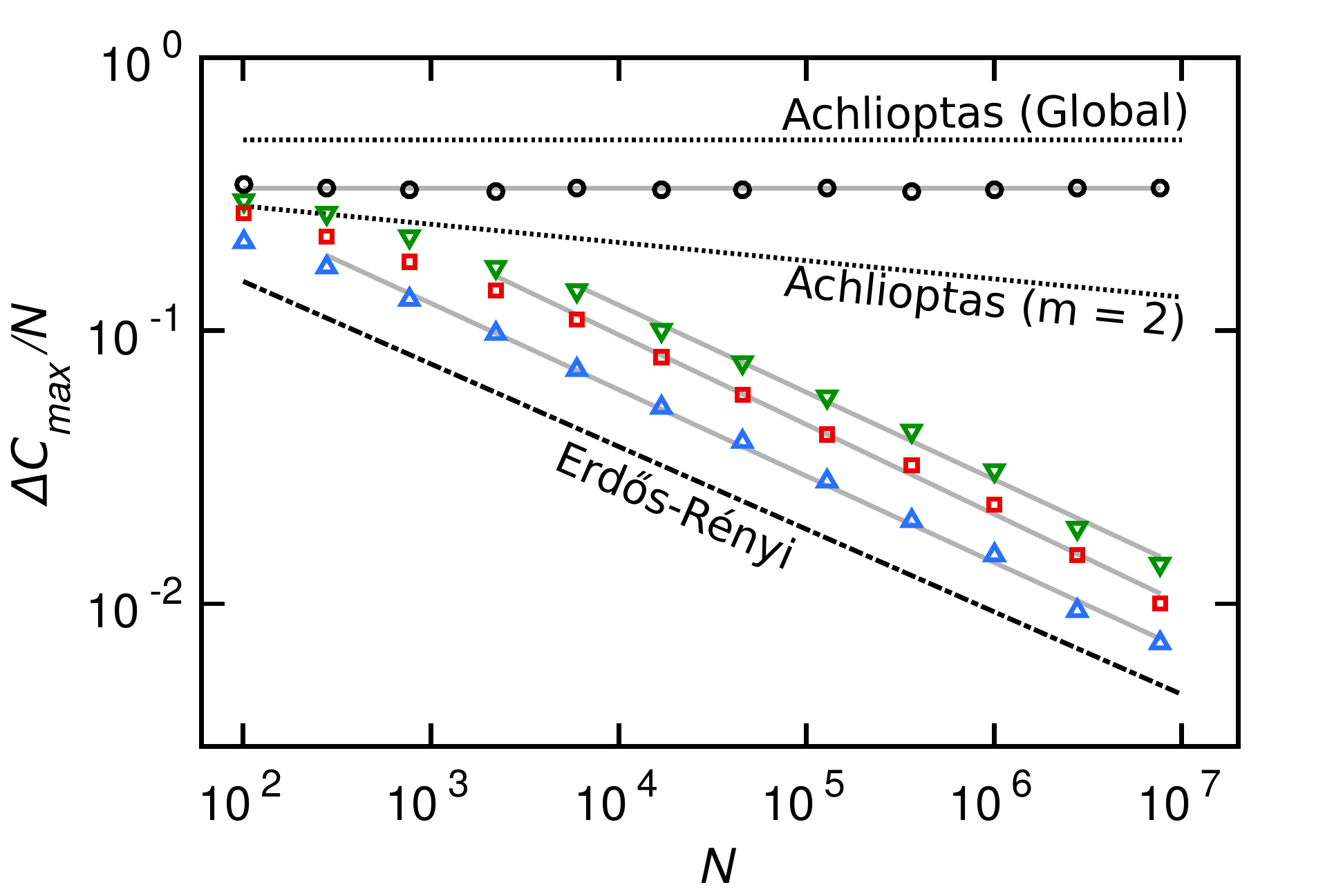}
  \caption{The average maximum jump in the order parameter as a function of system size for the DPR process with two choices (blue upward triangles), ten choices (red squares), fifty choices (green downward triangles), and global choice (black circles). Erd\H os-R\'{e}nyi (lower dashed line), as well as the AP with two choices (lower dotted line) and global choice (upper dotted line) are shown for comparison. Fits to the data (gray lines) for the three non-global DPR processes have decay exponents of $\omega = 0.316$, $\omega = 0.328$, and $\omega = 0.319$, respectively.}
  \label{MaxJumps}
\end{figure}

Increasing the number of choices does not appear to change the decay exponent in the DPR process, unlike in the AP \cite{singlelink}, which suggests that the locality of the information used in the DPR suppresses its ability to achieve the buildup of multiple large clusters that inevitably leads to bigger jumps in the order parameter. This is even more striking given that the value of the critical point increases from $p_{c} \approx 0.76$ for $m = 2$ choices to $p_{c} \approx 0.93$ for $m = 10$ choices and $p_{c} \approx 0.97$ for $m = 50$ choices, eventually asymptoting to $p_{c} \approx 1$ for global choice, implying that increasing the number of choices works to suppress the transition without actually building up the so-called ``powder keg'' conditions necessary to achieve explosiveness. Rather, the DPR works to constrict the degree distribution, as shown in Figure 3, which leads to something of a powder keg in the node degrees instead of cluster sizes. However, in contrast to explosive percolation, this degree-oriented powder keg does not ``ignite'' near the critical point.

Despite the lack of a ``powder keg,'' global choice in the DPR process nevertheless produces a first-order phase transition. We simulated global choice using the following process, as increasing the number of choices becomes computationally intensive at large system sizes. Initially, every node is randomly paired with another unpaired node, at which point the node pairs begin to join together and form chains. Only the two ends of each chain are candidates for edge addition, as they have degree $d = 1$ while internal nodes in the chain have degree $d = 2$. Eventually these chains will tend to form large, closed loops whenever the two ends of a single chain are randomly chosen to join together. The loops then merge together very close to $p = 1$, shortly after every node has degree $d = 2$, resulting in a critical point of $p_{c} \approx 1$ since the largest jump in the order parameter will tend to occur when two large loops merge. The result is a first-order phase transition with exclusively short-range information dictating its development. Similar to the AP with global choice, the largest jump for the DPR with global choice remains constant, with an approximate value of $\Delta C_{max}/N = 0.33$ for all $N$, shown in Figure 2. However, in the DPR process, the crossover from second-order to first-order appears to happen via the extension of a shoulder at increasing system sizes as the number of choices increases, rather than the typical rise in the slope of the power law seen in explosive percolation. Essentially, what appears to be finite-size effects observed with increasing number of choices could in fact be a signifier of a slow crossover to a discontinuous transition.

\begin{figure}
  \includegraphics[width=\linewidth]{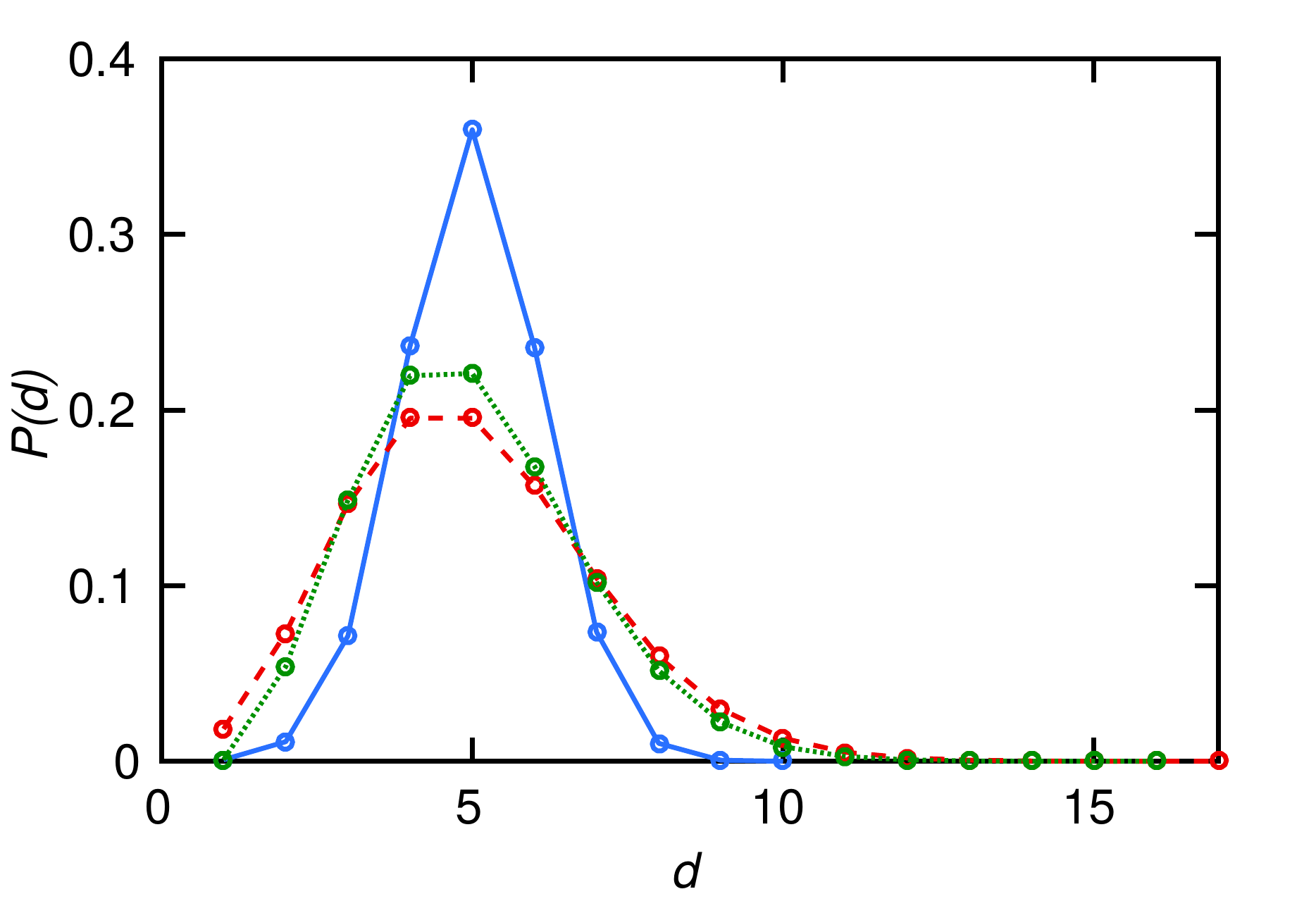}
  \caption{The degree distributions at $p = 5$ for Erd\H os-R\'{e}nyi (dashed red), degree product rule process (solid blue), and Achlioptas process (dotted green) at $N = 1.7\times10^{4}$ nodes and $m = 2$ choices for the DPR process and AP.}
  \label{DegreeDists}
\end{figure}

\textit{Criticality and universality.}--Second-order phase transitions are characterized by critical behavior which permits the use of scaling theory in determining universal behavior near the critical point \cite{FSS,numericalEP}. These functional forms are a result of the fact that all state variables associated with the phase transition behave as power laws near the critical point due to scale independence within the system. Using this process, one finds a rescaling of the order parameter for system size that has the following general form:

\begin{equation}
C = N^{-\beta/\nu}F[(p-p_c)N^{1/\nu}]
\end{equation}

The value of $\beta$ is associated with the behavior of the order parameter with system size, while $\nu$ scales the correlation length (mean distance between nodes in a cluster) with the distance to the critical point. The function $F$ is a universal function that allows collapse onto a single master curve. The average cluster size $S$ should rescale in a similar manner, though with a different critical exponent affecting the system size and a separate universal function $H$:

\begin{equation}
S = N^{\gamma/\nu}H[(p-p_c)N^{1/\nu}]
\end{equation}

\begin{table}[t]
\caption{Critical point $p_{c}$, and summary of critical exponents for the three growth processes discussed in this paper with $m=2$ for the AP and DPR processes.}
\centering
\label{t:CritExps}
\begin{tabularx}{\columnwidth}{Xsssss}
\noalign{\smallskip} \hline \hline \noalign{\smallskip}
Growth process & $p_{c}$ & $\beta/\nu$ & $\gamma/\nu$ & $\tau$ & $\omega$ \\
\hline
Erd\H os-R\'{e}nyi & 0.5 & 0.33 & 0.34 & 2.5 & 0.3 \\
\textbf{DPR} & \textbf{0.763} & \textbf{0.33} & \textbf{0.37} & \textbf{2.45} & \textbf{0.32} \\
Achlioptas & 0.888 & 0.02 & 0.48 & 2.08 & 0.065  \\
\noalign{\smallskip} \hline \noalign{\smallskip}
\end{tabularx}
\end{table}

\begin{figure}
  \includegraphics[width=\linewidth]{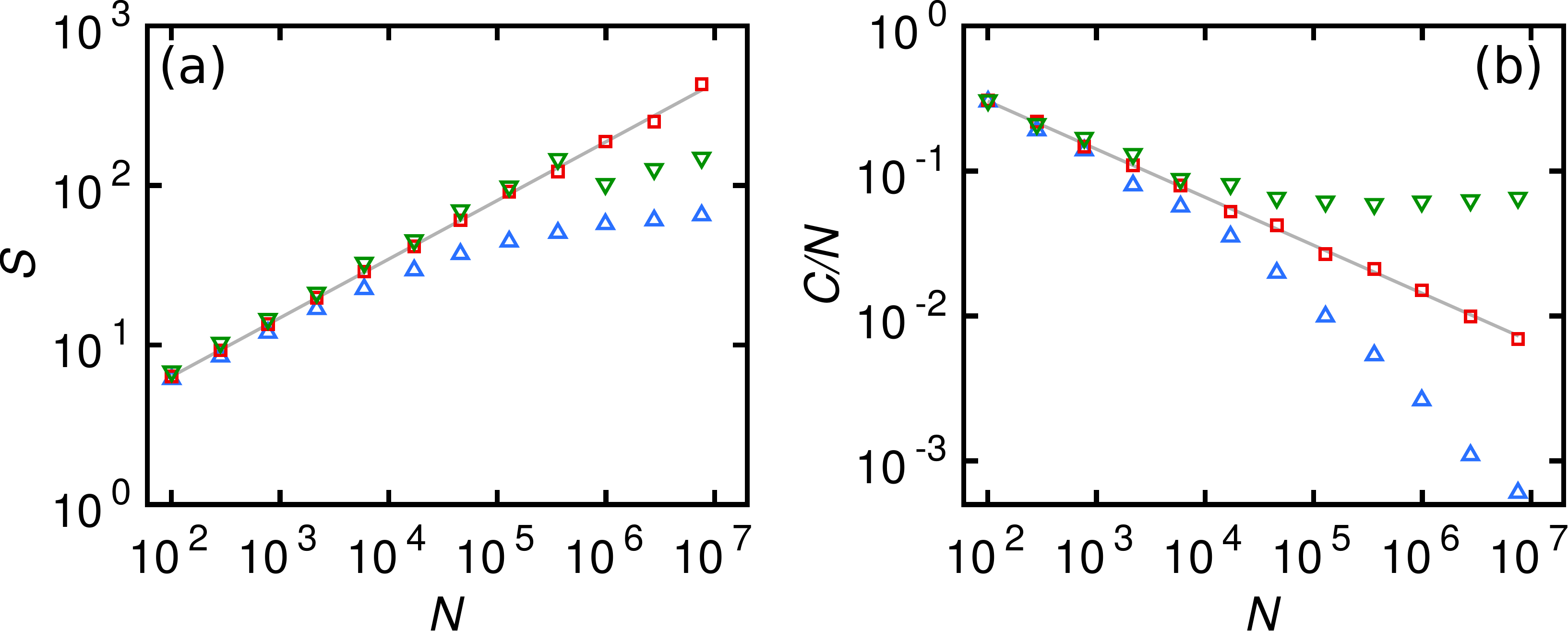}
  \caption{Finite-size scaling for the critical exponents $\beta/\nu$ and $\gamma/\nu$ of the DPR process. (a) Mean cluster size $S$ is plotted versus system size $N$. The fit at $p = p_{c} = 0.763$ (red squares) gives the value $\gamma/\nu = 0.37$. (b) Relative size of the largest cluster $C/N$ is plotted versus system size $N$. The fit at $p = p_{c} = 0.763$ (red squares) gives the value $\beta/\nu = 0.33$. Breakdown of the power law scaling away from the critical point is shown in both (a) and (b) for $p = 0.75$ (blue upward triangles) and $p = 0.77$ (green downward triangles).}
  \label{fig:CriticalExponents}
\end{figure}

Here, the exponent $\gamma$ scales the average cluster size (excluding the giant component) with system size $N$. Together, equations (1) and (2) contain the set of critical exponents and scaling functions required to characterize the DPR phase transition and allow universal collapse onto master curves. Measuring the critical exponents necessitates finding both the largest cluster size and the average size of clusters (excluding the largest) at the critical point for varying system sizes. The critical point serves as a separatrix for the largest cluster size--at the critical point it will follow a power law with growing system size, while above and below the critical point it will increasingly curve away from the separating line due to the excess (or deficit) of edges interrupting the scale-free nature of the system. The average size of the remaining clusters, however, will decay with growing system size both above and below the critical point due to the largest cluster absorbing an increasing portion of the nodes above the critical point. Figure 4 illustrates this behavior, which provides an additional check on the approximate value of the critical point, $p_{c} = 0.763$. The fits in Figure 4a and 4b provide values of $\beta/\nu = 0.33$ and $\gamma/\nu = 0.37$, respectively, for the scaling exponents of the DPR process. Again, these values draw comparisons to the classical Erd\H os-R\'{e}nyi process despite the fundamental differences in reversibility and information loss between the two growth processes.

\begin{figure}
  \includegraphics[width=\linewidth]{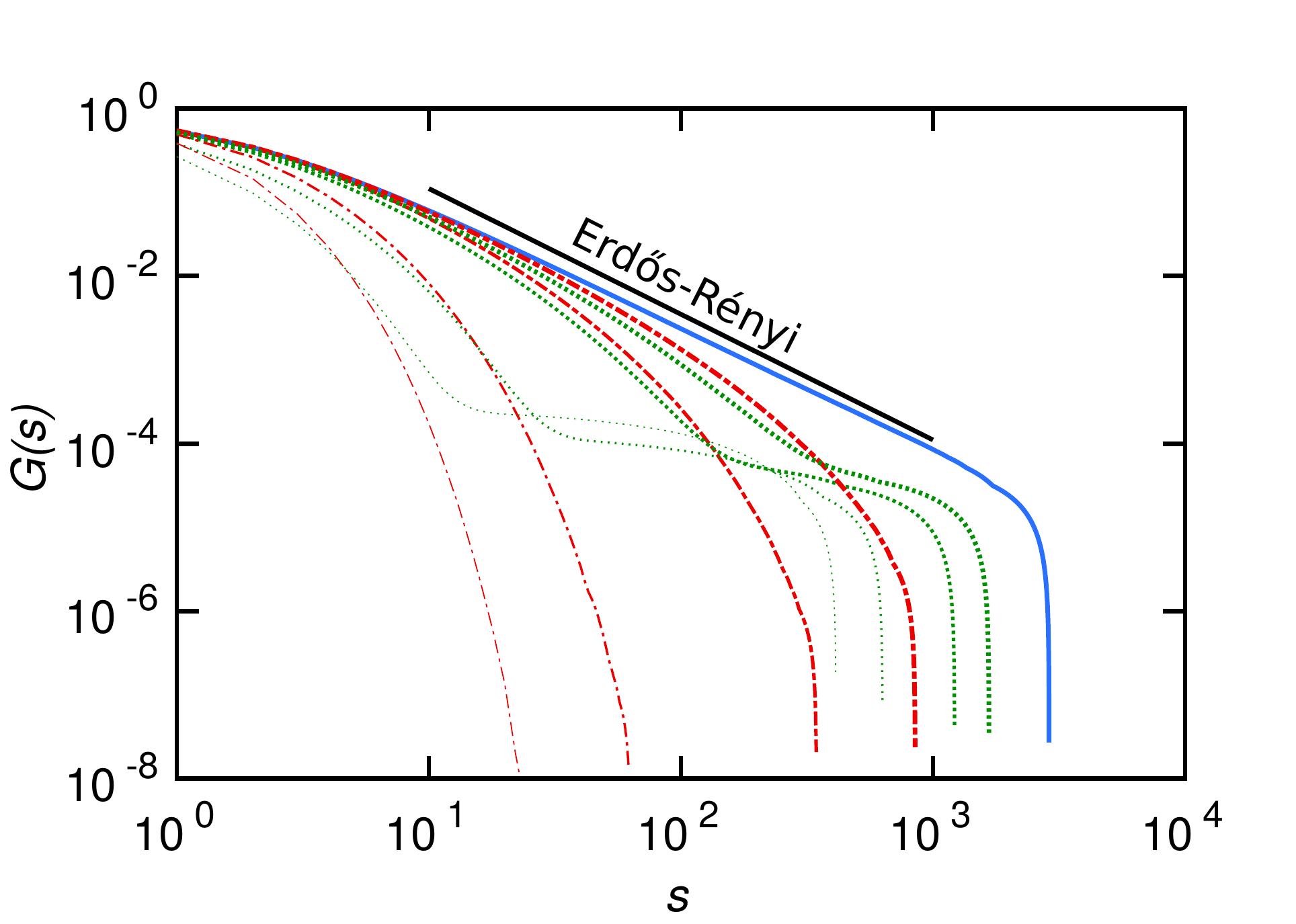}
  \caption{Cumulative distribution of cluster sizes at the critical point (solid blue) and at points above (green dotted) and below (red dot-dashed) the critical point for $N = 1.3\times10^{5}$ nodes. Thicker lines are nearer to the critical point. The solid black line is a guide for Erd\H os-R\'{e}nyi $(\tau = 2.5)$. The Fisher exponent, $\tau = 2.45$, is found by fitting a power law to the distribution at the critical point. Red dot-dashed curves are for $p = 0.38, 0.46, 0.54, 0.62$, green dotted curves are for $p = 0.92, 1, 1.08, 1.15$.}
  \label{fig:ClusterDists}
\end{figure}

Along with the set of critical exponents, the Fisher exponent $\tau$, which describes the power law decay of the cluster size distribution at the critical point, completes the picture of how the network percolates. By revealing the structure of cluster sizes beyond the largest component, the Fisher exponent provides details about how susceptible the network is to forming larger clusters near the critical point. Shown in Figure 5, the cluster size distribution at the critical point follows the form $G(s) \sim s^{1-\tau}$. The decay in cluster size for the DPR process is well-fit by a power law with $\tau = 2.45$, which may be consistent with Erd\H os-R\'{e}nyi $(\tau = 2.5)$. The cluster size distributions of the DPR process above the critical point show a mixture of explosive and classical behavior--plateaus form as in explosive percolation, however the distributions above the critical point remain entirely below the distribution at the critical point, as is the case in Erd\H os-R\'{e}nyi growth \cite{continuousEP}. This seems to suggest that the DPR process preferentially builds a few large clusters after the critical point, though it substantially delays building up the remaining smaller clusters as compared to explosive percolation. A comparison of the three growth processes considered in this paper is presented in Table 1.

\textit{Conclusions.}--Prescriptive processes for network growth, such as the one we presented, that tune percolation while circumventing the formation of a powder keg are useful in cases where connectivity is a liability. Here, we have described a way in which networks can be designed and grown that delays the onset of percolation without the risk of sudden connectivity, allowing for more manageable failure modes in cases where connectivity is undesirable. This growth scheme provides a new set of tools for researchers in a wide array of fields to use when intervening on growing networks, requiring a great deal less information when making decisions about how to guide networks towards more desirable topologies. In cases where acting quickly on a developing network is crucial, the DPR can be enacted with ease whereas enacting cluster-oriented growth schemes may be impractical.

Our work establishes the fact that in order to turn a percolation transition from second-order to first-order one need not necessarily have access to global information, as in explosive percolation. In addition, the use of local information extends the lower bound for explosive percolation to even lower critical connectivities than previously accessible with global information.

The selection criteria in DPR grown networks could be further altered in order to use the product of degrees of second-nearest, or third-nearest neighbors, etc., methodically extending the distance with which information about connectivity is communicated within a network. Such a tool could allow for improved modeling of networks where interactions extend to a finite distance. Degree rule processes may also be of interest within the context of core percolation \cite{coreperc}, as they naturally produce networks with larger cores due to the narrow width of the degree distribution compared to traditional and explosive percolation.

\begin{acknowledgments}
This work was supported by the NSF under Career Grant No. DMR-1255370 and a grant from the Simons Foundation No. 454939. The ACISS supercomputer is supported under a Major Research Instrumentation grant, Office of Cyber Infrastructure, No. OCI-0960354.

A.J.T., G.T., and E.I.C. contibuted equally to this work.
\end{acknowledgments}

\end{document}